\documentclass[a4paper,11pt]{article}
\pdfoutput=1 % if your are submitting a pdflatex (i.e. if you have
\usepackage{jheppub} % for details on the use of the package, please
\usepackage[T1]{fontenc} % if needed

\usepackage{amssymb}
\usepackage{amsthm}
\usepackage{xspace}
\usepackage{bookmark}
\usepackage{tabstackengine}
\usepackage{todonotes}
\usepackage{subfig}
\allowdisplaybreaks

\def\ep  {\ensuremath{\epsilon}}
\def\J {{\boldsymbol J}}
\def\c {{\boldsymbol c}}
\def\T {{T}}
\def\L {{L}}

\def\OO {\ensuremath{O}}
\newcommand{\bs}{\boldsymbol}

\newcommand{\LiteRed}{\texttt{LiteRed}\xspace}
\newcommand{\Libra}{\texttt{Libra}\xspace}

\title{Master integrals for bipartite cuts of three-loop photon self energy.}

\author[a]{ R.N. Lee}
\author[b,a,c]{A.I. Onishchenko}

\affiliation[a]{Theory department, Budker Institute of Nuclear Physics,\\ Novosibirsk, Russia}
\affiliation[b]{Bogoliubov Laboratory of Theoretical Physics, Joint
	Institute for Nuclear Research,\\ Dubna, Russia}
\affiliation[c]{Skobeltsyn Institute of Nuclear Physics, Moscow State University\\ Moscow, Russia}

\emailAdd{r.n.lee@inp.nsk.su}
\emailAdd{onish@theor.jinr.ru}

\usepackage{blindtext}

\abstract{
    We calculate master integrals for bipartite cuts of the three-loop propagator QED diagrams. These master integrals determine the spectral density of the photon self energy. Our results are expressed in terms of the iterated integrals, which, apart from the $4m$ cut, reduce to Goncharov's polylogarithms. The master integrals for $4m$ cut have been calculated in our previous paper in terms of the one-fold integrals of harmonic polylogarithms and complete elliptic integrals. We provide the threshold and high-energy asymptotics of the master integrals found, including those for $4m$ cut. 
}

\begin{document} 
\maketitle
\flushbottom
\section{Introduction}

Photon self-energy operator $\Pi(s)=\sum_n \left(\frac{\alpha}{\pi}\right)^n \Pi_n(s)$ is a fundamental physical quantity of quantum electrodynamics. It is an important ingredient of many physically relevant calculations. One-loop result $\Pi_1(s)$ is presented in many QED textbooks (see, e.g., \cite{berestetskii1982quantum}), and the two-loop contribution $\Pi_2(s)$ has been calculated long ago by K\"allen and Sabry in Ref. \cite{kallen1955k}. 
The three-loop contribution also appears in many applications, see, in particular, Refs. \cite{Baikov:1995ui,kinoshita1999sixth,kinoshita1999accuracy}. Some calculations require knowing this object with very high precision.  Kinoshita and Lindquist have derived 6-fold parametric representation for the renormalized three-loop photon self-energy in a dedicated work  \cite{kinoshita1983parametric}. Except for this computationally quite expensive representation, no exact analytical representation for $\Pi_3(s)$ has been derived so far. Baikov and Broadhurst \cite{Baikov:1995ui} have derived a simple Pad\`e approximation  for this operator using a few terms of the asymptotic expansions near 3 special points: $s=0,4m^2,\infty$ ($m$ is the electron mass). The precision of this approximation has been estimated indirectly, by comparing the contribution of the 3-loop polarization operator to 4-loop $g_\mu-2$. The conservative estimate of this precision given by Baikov and Broadhurst was as high as $0.002$ percent for this case. This declared precision is quite remarkable and it would be interesting to compare the exact result with the approximate one in a more direct way. In particular, we are concerned about the impact of the second threshold $s=16m^2$ which becomes relevant starting from three loops.

\begin{figure}
    \includegraphics[width=1\textwidth]{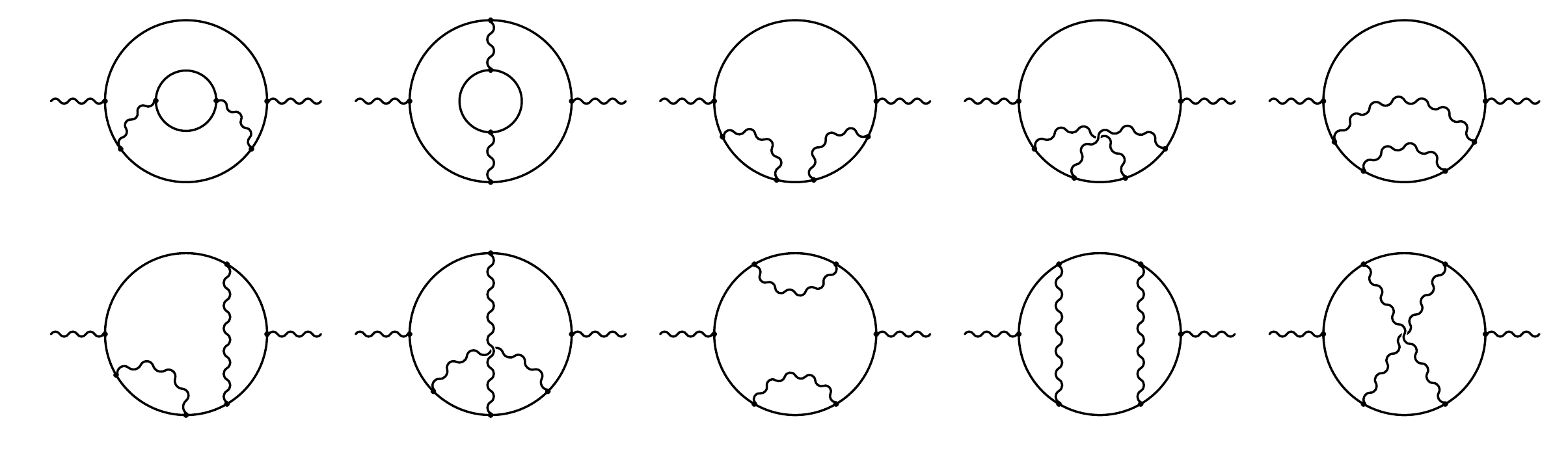}
    \caption{Diagrams contributing to $3$-loop photon self energy in QED.}
    \label{fig:diagrams}
\end{figure}

In the present paper we make an important step towards analytic  calculation of the 3-loop polarization operator. Namely, we derive the analytic expressions for all master integrals which are required for the calculation of the spectral density $\rho(s)=\Im \Pi_3(s+i0)/\pi$. Thanks to Cutkosky rules, this spectral density is expressed via bipartite cuts of a specific set of three-loop massive diagrams depicted in Fig. \ref{fig:diagrams}.

Classifying the cuts by the number of massive and massless lines that are cut, we have $(2,0)$, $(2,1)$, $(2,2)$, and $(4,0)$ cuts, where $(k,l)$ denotes the cut with $k$ massive and $l$ massless lines being cut. The non-polylogarithmic integrals appear only in $(4,0)$ cuts due to the presence of three-loop cut banana graph among the master integrals. Fortunately, all $(4,0)$ cuts are free from divergences of both ultraviolet and infrared origin, and, therefore, can be calculated exactly in $d=4$. In Ref. \cite{Lee:2019wwn} we have presented a set of the master integrals 
sufficient for this purpose. The master integrals have been expressed via 
iterated integrals with all weights but one being the rational functions. We 
have presented also the expressions in terms of one-fold integrals with 
integrands expressed via complete elliptic integrals and generalized 
polylogarithms. 

We calculate below the master integrals for the remaining cuts $(2,0)$, $(2,1)$, and $(2,2)$. We express the exact results for these cuts via iterated integrals with all weights being rational. These iterated integrals are, therefore, expressible in terms of Goncharov's polylogarithms. 
Besides he application to the photon self-energy, these master integrals enter the total cross sections of one-photon electron-positron annihilation to muons or to hadrons at NNLO.

For the $(4,0)$ cut we write the results of Ref.  \cite{Lee:2019wwn}  in terms of similar iterated integrals, with the right-most weight expressed via complete elliptic integrals. We provide explicit expressions for the asymptotics of the obtained integrals, including those from $(4,0)$ class, near thresholds and in the high-energy limit. 

Our calculation follows the standard path:
\begin{enumerate}
    \item IBP reduction. Constructing differential equations for master integrals.
    \item Reduction of the differential equations to $\epsilon$-form.
    \item Fixing boundary conditions from threshold asymptotics.
    \item Constructing solution in terms of iterated integrals.
\end{enumerate}

\section{Prototypes, Laporta bases and boundary conditions}

As it was already mentioned in the Introduction, our calculation strategy for polylogarithmic master integrals is based on the reduction of a differential system to $\ep$-form \cite{epform1,epform2}. In the case of finite non-polylogarithmic master integrals we use instead the notion of $\ep$-regular basis \cite{Lee:2019wwn}. In what follows we will use $\bs  j$ for the column of Laporta master integrals and $\bs  J$ for the canonical or $\ep$-regular basis.

Having reduced the system of differential equations for polylogarithmic master integrals to $\ep$-form the latter can be easily solved in terms of multiple polylogarithms and the solution for initial Laporta master integrals can be written as
\begin{equation}
\J_{\rm Laporta}(\beta) = e^{-3\ep \gamma_E}\T(\beta)\J_{\rm canonical}(\beta) = \T(\beta) \mathrm{Pexp}\left[
\ep \int_0^{\beta} S(t) dt
\right]\L\cdot \c e^{-3\ep \gamma_E}\, , \label{canonical-system}
\end{equation}
where $\beta = \sqrt{1-4/s}$, $\T$ is the transformation matrix to the canonical basis, and $S$ is the matrix entering differential equations system for the canonical master integrals
\begin{equation}
\partial_{\beta}\J_{\rm canonical}(\beta) = \ep S (\beta)\J_{\rm canonical}(\beta), 
\end{equation} 
$\c$ is the column of the coefficients in threshold asymptotic expansions of the Laporta master integrals, and $\L$ is a rational matrix depending on $\ep$.
 
The reduction to $\ep$-form, the choice of the coefficients $\c$, and the determination of the corresponding ``adapter'' matrix $\L$ is made with the help of \Libra package \cite{lee2020libra}.

%boundary constants of the form $c_i(\beta^{a+b\ep})$, where $c_i (\beta^{a+b\ep})$ is the constant in front of $\beta^{a+b\ep}$ in the threshold expansion of $i$-th Laporta master integral. The index $i$ does not necessary take all the values from the set $\{1,\ldots \dim\J_{\rm Laporta}(\beta)\}$ and one can have several constants defined for the same master integral $i$, such that the total number of constants is equal to the number of master integrals.
%
%The exact in $\ep$ matrix $\L$ required to use vector of constants $\c$ in boundary conditions can be easily determined with the use of Frobenius solution of system \eqref{canonical-system} instead of $P$-exponent. 

It appears that all required families of integrals can be conveniently described using the following nine ``denominators'':
\begin{gather}\nonumber
    D_1=1-l_2^2,\ \
    D_2=1-l_3^2,\ \ 
    D_3=-\left(l_2-l_3\right)^2,\ \ 
    D_4=-\left(l_1-l_2\right)^2,\ \ 
    D_5=1-l_1^2,\\
    D_6=1-\left(q-l_3\right)^2\!\!,\  
    D_7=1-\left(q-l_1\right)^2\!\!,\ 
    D_8=1-\left(q-l_1+l_2-l_3\right)^2\!\!,\ 
    D_9=1-\left(q-l_2\right)^2\!\!.
    \label{eq:Ds}
\end{gather}

\subsection*{$(2,0)$ cuts}

\begin{figure}
    \centering
    \includegraphics[width=1\textwidth]{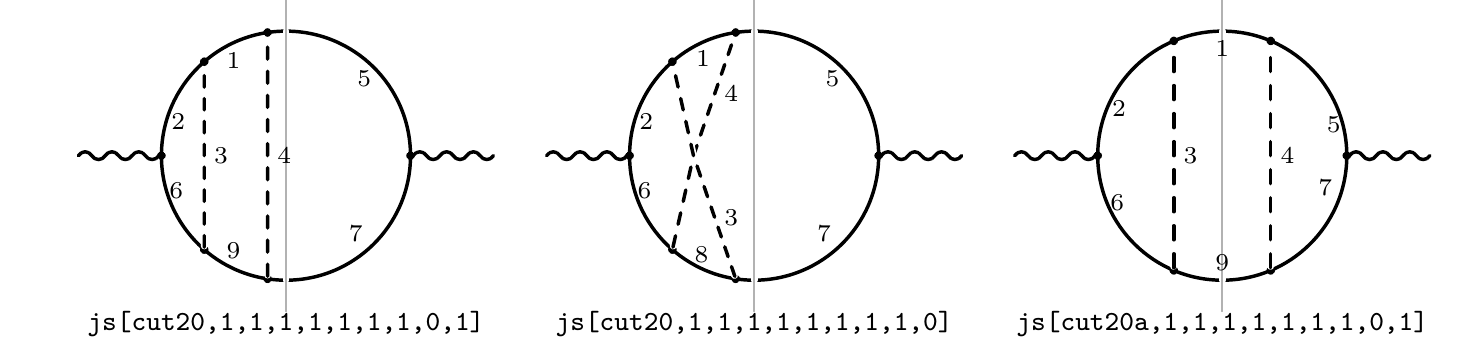}
    \caption{Prototypes for master integrals of $(2,0)$ cuts.}
    \label{fig:cut2m-protos}
\end{figure}

In the case of $(2,0)$ cuts all Laporta master integrals belong to one of the three prototypes shown in Fig. \ref{fig:cut2m-protos}. As the two first prototypes differ only by one denominator, we use one \LiteRed{} basis for them.\footnote{For the IBP reduction we use \LiteRed package, Ref. \cite{Lee2013a,Lee2021}.} Namely, we have
\begin{align}
j^{\texttt{cut20}}_{n_1\ldots,n_9} &= \frac{(2\pi)^2}{2\pi^{d/2}}\int\frac{d^dl_1 d^dl_2 d^dl_3}{(i\pi^{d/2})^2}\frac{\delta^{(n_5-1)}(-D_5)\delta^{(n_7-1)}(-D_7)}{\prod_{k=1,k\not\in \{5,7\}}^{9} (D_k-i0)^{n_k}} \\ 
j^{\texttt{cut20a}}_{n_1\ldots,n_9} &= \frac{(2\pi)^2}{2\pi^{d/2}}\int\frac{d^dl_1 d^dl_2 d^dl_3}{(i\pi^{d/2})(-i\pi^{d/2})}\frac{\delta^{(n_1-1)}(-D_1)\delta^{(n_9-1)}(-D_9)}{\prod_{k=1,k\not\in \{1,9\}}^{9} (\sigma_kD_k-i0)^{n_k}}\,,
\end{align}
where one of $n_8, n_9$ is necessarily non-positive in the first family and $n_8\leqslant 0$ in the second. We also use notation $\sigma_k=-1$ if $k\in\{4,5,7\}$ and $\sigma_k=1$ otherwise.
The denominators are defined in Eq.  \eqref{eq:Ds}.

We have 17 master integrals in the first family and 2 additional master integrals in the second:
\begin{center}
    \includegraphics[width=\textwidth]{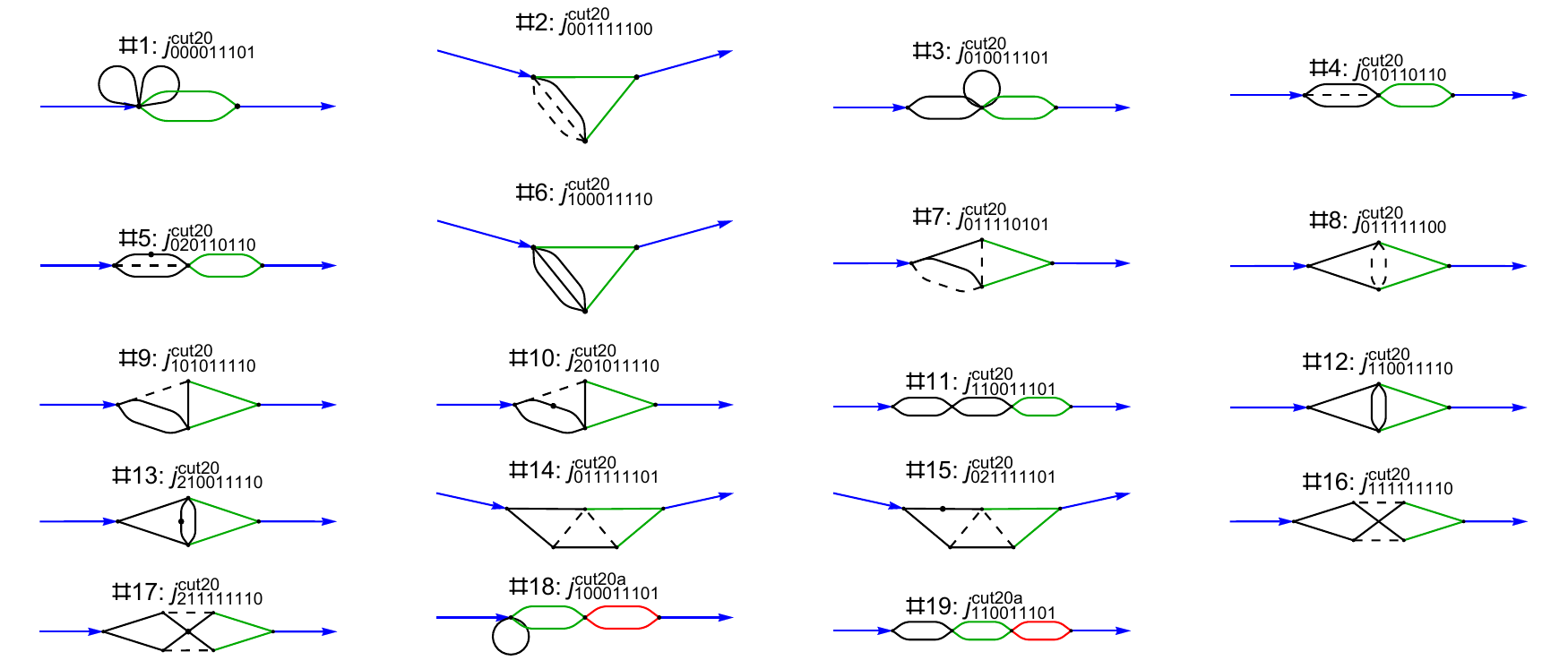}
\end{center}
Here the green lines correspond to the cut propagators, while the red lines correspond to complex conjugated part of the diagram. Note that the integral $j_{100011101}^{\texttt{cut20a}}$ is a complex conjugate of $j_{010011101}^{\texttt{cut20}}$ and we introduce it because it enters the differential equation for $j_{110011101}^{\texttt{cut20a}}$.

Using \Libra we have chosen the following constants to fix our boundary conditions:
\begin{multline}
\c =(
c_1(\beta^{1-2\ep}),\,  c_2(\beta^{1-2\ep}),\,  c_3(\beta^{2-4\ep}),\,  c_4(\beta^{6-8\ep}),\,  c_4(\beta^{1-2\ep}),\,  c_6(\beta^{1-2\ep}),\,  c_7(\beta^0),\, \,\nonumber \\ c_8(\beta^{2-6\ep})\, , c_9(\beta^0), c_9(\beta^{3-6\ep}),\, c_{11}(\beta^{3-6\ep}),\, c_{12}(\beta^0),\, c_{12}(\beta^{1-4\ep}),\, c_{14}(\beta^0),\,\nonumber \\ c_{14}(\beta^{1-6\ep}),\, c_{16}(\beta^0), \, c_{16}(\beta^{-1-6\ep}),\, c_{18}(\beta^{2-4\ep}),\, c_{19}(\beta^{3-6\ep})
)^\intercal,
\end{multline}
where  
$c_i (\beta^{\nu})$ 
denotes the coefficient in front of $\beta^{\nu}$ in the threshold asymptotics of $i$-th Laporta master integral. We calculate the required integrals using the expansion-by-regions method, Ref. \cite{Beneke:1997zp}. 

Let us present a few examples of calculation of the required coefficients. First, note that the above master integrals represent either the 2-loop vertex or the product of 1-loop vertices integrated over 2-particle phase space. The later integration decouples completely and we have
\begin{equation}
\Phi_2 = \frac{(2\pi)^2}{2}\int\frac{d^dl_1}{\pi^{d/2}}\delta (l_1^2-1)\delta ((q-l_1)^2-1) = \frac{\pi^{3/2}}{2\Gamma (3/2-\ep)}\left(\frac{s}{4}\right)^{-\ep}\beta^{1-2\ep}
\end{equation}
Noting that it scales as $\sim\beta^{1-2\ep}$ we immediately conclude that
\begin{equation}
c_7(\beta^0) = c_9(\beta^0) = c_{12}(\beta^0) = c_{14}(\beta^0) = c_{16}(\beta^0) = 0\, .
\end{equation}
Next, a careful inspection of expansion regions at the threshold \cite{threshold-expansion} gives
\begin{equation}
c_{14}(\beta^{1-4\ep}) = c_{16}(\beta^{-1-6\ep}) = 0\, .
\end{equation}
Let us now present some details of calculation of  $c_4(\beta^{1-2\ep})$ constant, which appears to be expressible via hypergeometric function $\,_{3}F_{2}$. We use the alpha parametrization for 2-loop sunset subdiagram and obtain
\begin{align}
    c_4(\beta^{1-2\ep}) &= \frac{\pi^{3/2}\Gamma (-1+2\ep)}{2\Gamma (3/2-\ep)} \int_0^{\infty}d\alpha_2\int_0^{\infty}d\alpha_3\int_0^{\infty}d\alpha_6 \delta(1-\alpha_2-\alpha_3-\alpha_6)\\
    &\times\left(\alpha _6 \left(\alpha _2-\alpha _3\right)^2+\alpha _2 \alpha _3 \left(\alpha _2+\alpha _3\right)\right)^{1-2 \epsilon } \left(\alpha _3 \alpha _6+\alpha _2 \left(\alpha _3+\alpha _6\right)\right)^{3 (\epsilon -1)}.
\end{align}
Using Cheng-Wu theorem \cite{ChengWu} we replace $\delta (1-\alpha_2-\alpha_3-\alpha_6)$ with $\delta (1-\alpha_2-\alpha_3)$ and making the change of variables $\alpha_2=x,\ \alpha_3=1-x,\ \alpha_6=x(1-x) y$ we obtain
\begin{equation}
   c_4(\beta^{1-2\ep})= \frac{\pi^{3/2}\Gamma (-1+2\ep)}{2\Gamma (3/2-\ep)} \int_0^{1}dx\int_0^{\infty}dy\,( x(1-x))^{\epsilon -1} (y+1)^{3 \epsilon -3} \left((1-2 x)^2 y+1\right)^{1-2 \epsilon }
\end{equation}
Now we use the Mellin-Barnes parametrization
\begin{equation}
    \frac{\Gamma (-1+2\ep)}{\left((1-2 x)^2 y+1\right)^{2 \epsilon-1}}=\intop_{-i\infty}^{i\infty} \frac{d\tau}{2\pi i}  \frac{\Gamma(-1+2\ep-\tau)\Gamma(\tau)e^{i\pi \tau} }{(1 + y)^{-1+2\ep -\tau}(4y x(1-x))^{\tau}}
\end{equation}
After this the integrals over $x$ and $y$ can be taken in terms of $\Gamma$-functions and we obtain
one-fold Mellin-Barnes representation
\begin{equation}
    c_4(\beta^{1-2\ep})= \frac{\pi^{5/2}}{2\Gamma (3/2-\ep)}\intop_{-i\infty}^{i\infty} \frac{d\tau}{2\pi i} \frac{ 4^{-\tau } e^{i \pi \tau } \Gamma (1-\epsilon ) \Gamma (\epsilon -\tau )^2 \Gamma (2 \epsilon -\tau -1)}{\sin(\pi  \tau ) \Gamma (2 \epsilon -2 \tau ) \Gamma (-\epsilon -\tau +2)}
\end{equation}
We close the contour of integration to the left and finally obtain
\begin{equation}
     c_4(\beta^{1-2\ep}) = \frac{\pi ^{5/2} 2^{2-4 \epsilon } \Gamma (2 \epsilon -1) }{\sin (\pi  \epsilon ) \Gamma (3-2 \epsilon ) \Gamma \left(\epsilon +\frac{1}{2}\right)}\, _3F_2\left(1,\epsilon ,2 \epsilon -1;2-\epsilon ,\epsilon +\tfrac{1}{2};1\right).
\end{equation}

In a similar way, we have been able to fix all required boundary constants exactly in $\ep$, with two of them expressed via hypergeometric functions.
E.g., we have
\begin{equation}
   c_6(\beta^{1-2\ep}) =  \frac{\pi ^2  \Gamma (\epsilon ) \left(\, _3F_2\left(\frac{1}{2},1,2 \epsilon -1;2-\epsilon ,\epsilon +\frac{1}{2};1\right)+(1-2 \epsilon)\, _3F_2\left(1,\frac{3}{2}-\epsilon ,\epsilon ;\frac{3}{2},3-2 \epsilon ;1\right)\right)}{2^{2 \epsilon -1 } \Gamma (3-2 \epsilon )(2 \epsilon -1)\sin (\pi  \epsilon )}\,.
\end{equation}
We have checked that a few leading terms of $\ep$-expansion of $c_i^{a+b\ep}$ constants calculated here can be reproduced from the results of  2-loop vertex diagrams calculation presented in \cite{Bonciani:2003te}.

\subsection*{$(2,1)$ cuts}
\begin{figure}
    \centering
    \includegraphics[width=0.67\textwidth]{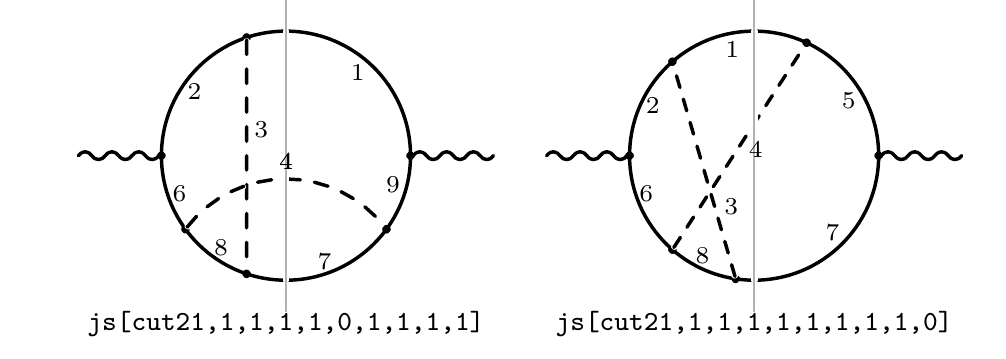}
    \caption{Prototypes for master integrals of $(2,1)$ cuts.}
    \label{fig:cut2m1z-protos}
\end{figure}

All Laporta master integrals for $(2,1)$ cuts belong to two prototypes in Fig. \ref{fig:cut2m1z-protos}. They are defined as
\begin{align}
    j^{\texttt{cut21}}_{n_1\ldots,n_9} &= \frac{(2\pi)^3}{2\left(\pi^{d/2}\right)^2}\int\frac{d^dl_1 d^dl_2 d^dl_3}{i\pi^{d/2}}\frac{\prod_{k\in \{1,4,7\}}\delta^{(n_k-1)}(-D_k)}{\prod_{k=1,k\not\in \{1,4,7\}}^{9} (D_k-i0)^{n_k}}\,,
\end{align}
where one of $n_5, n_9$ is necessarily non-positive and $D_k$ are defined in Eq. \eqref{eq:Ds}.
We have 22 master integrals:
\begin{center}
    \includegraphics[width=\textwidth]{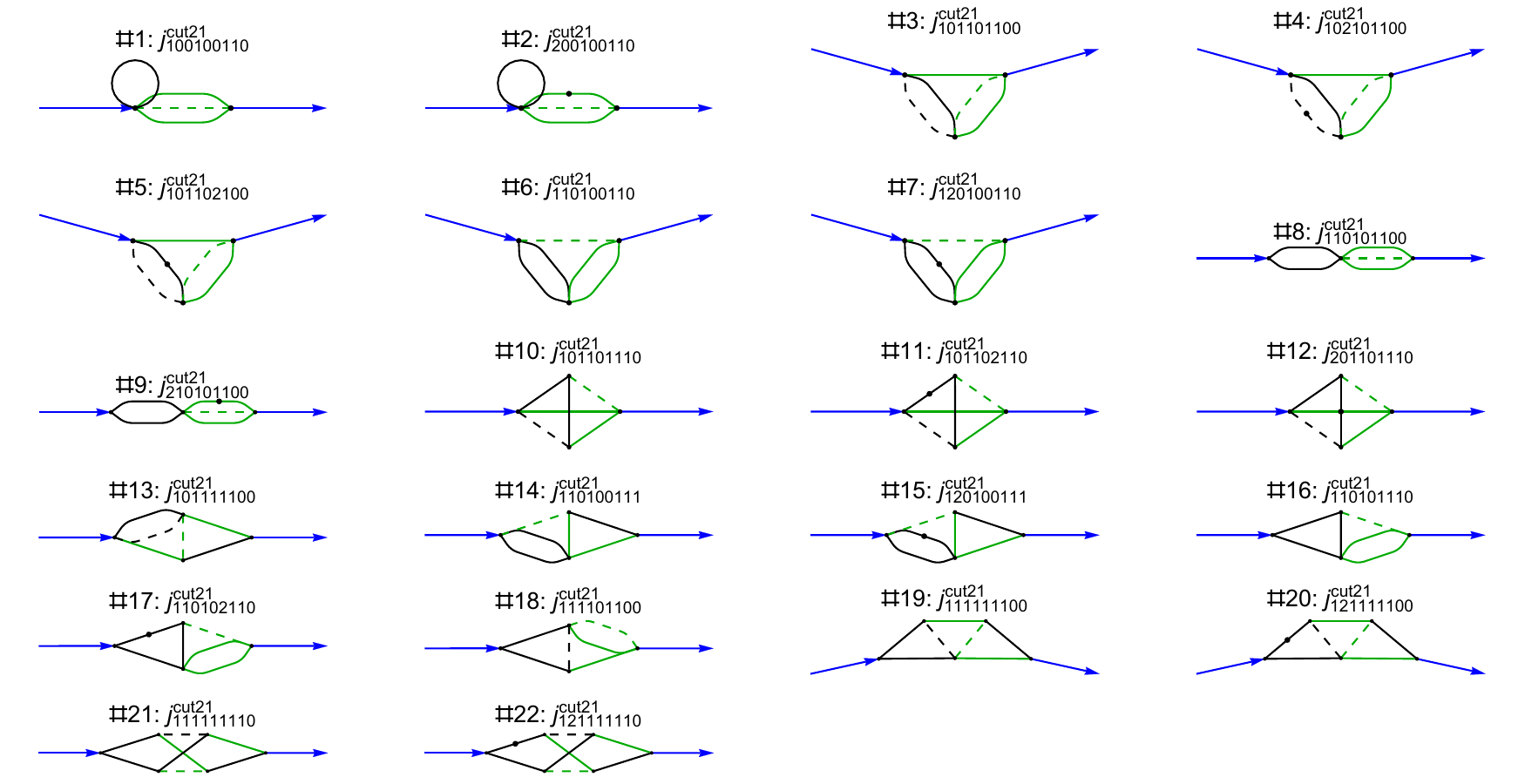}
\end{center}
We fix boundary conditions by calculating the following coefficients in the threshold asymptotics of the Laporta master integrals:

\begin{multline}
    \c = \Big(c_1(\beta^0),c_1\left(\beta ^{5-6 \epsilon }\right),c_3(\beta^0),c_3\left(\beta ^2\right),c_3\left(\beta ^{7-10 \epsilon }\right),c_6(\beta^0),c_6\left(\beta ^{6-8 \epsilon }\right),c_8\left(\beta ^{6-8 \epsilon }\right),\\
    c_8\left(\beta ^{1-2 \epsilon }\right),
    c_{10}(\beta^0),c_{10}\left(\beta ^2\right),c_{10}\left(\beta ^{3-6 \epsilon }\right),c_{13}(\beta^0),c_{14}(\beta^0),c_{15}\left(\beta ^{1-6 \epsilon }\right),\\
    c_{16}(\beta^0),c_{16}\left(\beta ^{3-6 \epsilon }\right)
    ,c_{18}(\beta^0),c_{19}(\beta^0),c_{20}\left(\beta ^{-6 \epsilon -1}\right),c_{21}(\beta^0),c_{21}\left(\beta ^{-6 \epsilon -1}\right)\Big)^\intercal\,.
\end{multline}
among which only four are nonzero:
\begin{gather}
c_1\left(\beta ^{5-6 \epsilon }\right)=\frac{\pi ^{5/2} 2^{1-2 \epsilon } \csc (\pi  \epsilon )}{(\epsilon -1) \Gamma \left(\frac{7}{2}-3 \epsilon \right)},\ c_3\left(\beta ^{7-10 \epsilon }\right)=\frac{\pi ^2 2^{1-2 \epsilon } e^{2 i \pi  \epsilon } \epsilon  \Gamma (3-4 \epsilon ) \Gamma (-\epsilon ) \Gamma (2 \epsilon -1)}{\Gamma \left(\frac{9}{2}-5 \epsilon \right) \Gamma \left(\frac{3}{2}-\epsilon \right)},\nonumber\\
c_6\left(\beta ^{6-8 \epsilon }\right)=\frac{ \pi ^{3/2} 2^{1-4 \epsilon } e^{i \pi  \epsilon } \Gamma (1-\epsilon )^2 \Gamma \left(\epsilon -\frac{1}{2}\right)}{i\Gamma (4-4 \epsilon )},\ c_8\left(\beta ^{6-8 \epsilon }\right)=\frac{\pi ^2 4^{-\epsilon } e^{i \pi  \epsilon } \Gamma (1-\epsilon ) \Gamma \left(\epsilon -\frac{1}{2}\right)}{i\Gamma \left(\frac{7}{2}-3 \epsilon \right)}\,.
\end{gather}

It is remarkable that the differential equations for $(2,1)$-cut master integrals are the only ones which have singularity at $s=1$. Consequently the $(2,1)$-cut master integrals can not be expressed via harmonic polylogarithms because they involve involve Goncharov's polylogarithms with letters $\pm\sqrt{3}i$.

\subsection*{$(2,2)$ cuts}

\begin{figure}
    \centering
    \includegraphics[width=1\textwidth]{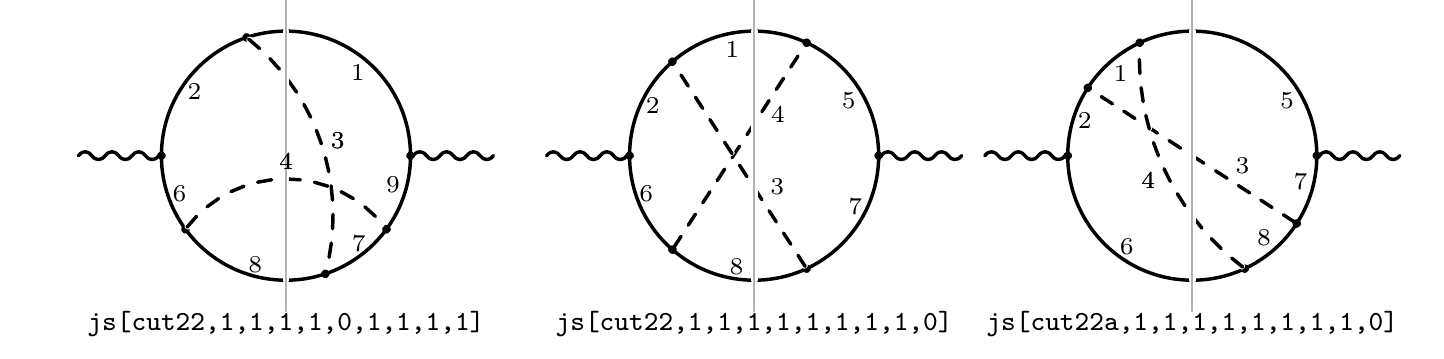}
    \caption{Prototypes for master integrals of $(2,2)$ cuts.}
    \label{fig:cut2m2z-protos}
\end{figure}

All Laporta master integrals for $(2,2)$ cuts belong to three prototypes in Fig. \ref{fig:cut2m2z-protos}. They are defined via two \LiteRed{} bases
\begin{align}
    j^{\texttt{cut22}}_{n_1\ldots,n_9} &= \frac{(2\pi)^4}{2\left(\pi^{d/2}\right)^3}\int d^dl_1 d^dl_2 d^dl_3\frac{\prod_{k\in \{1,3,4,8\}}\delta^{(n_k-1)}(-D_k)}{\prod_{k=1,k\not\in \{1,3,4,8\}}^{9} (D_k-i0)^{n_k}}\,,\\
    j^{\texttt{cut22a}}_{n_1\ldots,n_9} &= \frac{(2\pi)^4}{2\left(\pi^{d/2}\right)^3}\int d^dl_1 d^dl_2 d^dl_3\frac{\prod_{k\in \{3,4,5,6\}}\delta^{(n_k-1)}(-D_k)}{\prod_{k=1,k\not\in \{3,4,5,6\}}^{9} (D_k-i0)^{n_k}}\,,
\end{align}
where one of $n_5, n_9$ is necessarily non-positive for the first basis and $n_9\leqslant0$ for the second. The functions $D_k$ are defined in Eq. \eqref{eq:Ds}.
We have 15 master integrals:
\begin{center}
    \includegraphics[width=\textwidth]{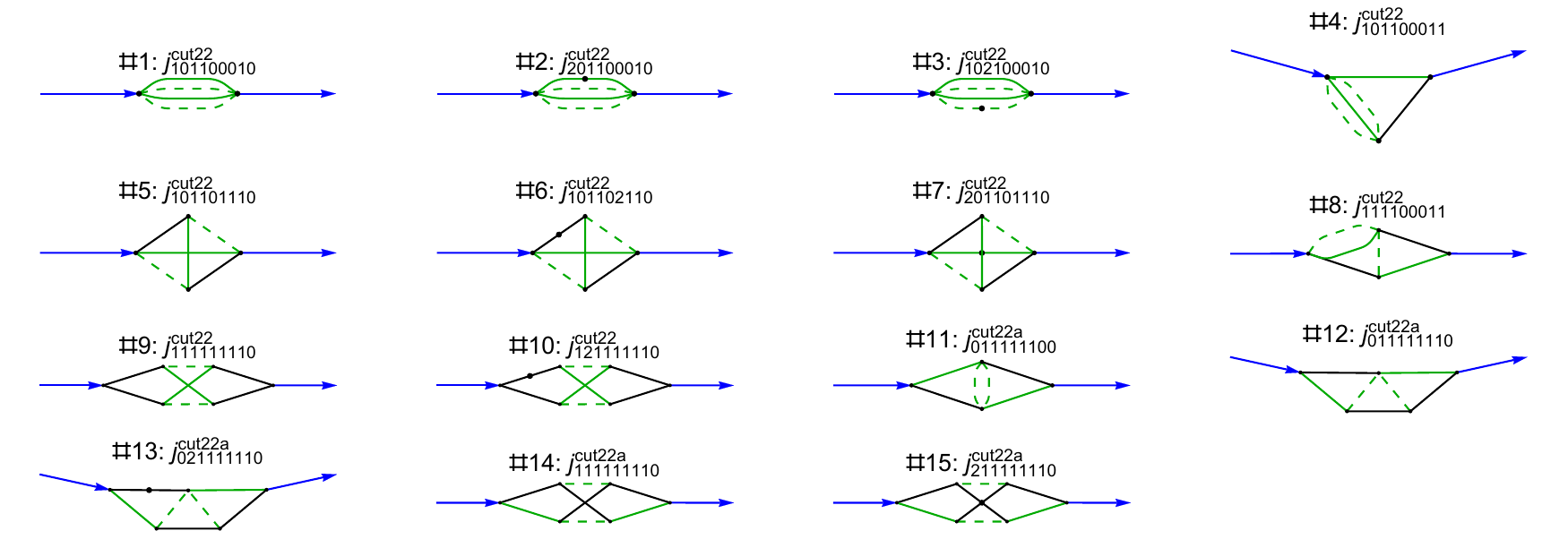}
\end{center}
We fix boundary conditions by calculating the following coefficients in the threshold asymptotics of the Laporta master integrals:

\begin{multline}
    \c = \Big(c_1(\beta^0),c_1\left(\beta ^2\right),c_1\left(\beta ^{9-10 \epsilon }\right),c_4\left(\beta ^{1-2 \epsilon }\right),c_5(\beta^0),c_5\left(\beta ^2\right),c_5\left(\beta ^{3-6 \epsilon }\right),c_8(\beta^0),c_9(\beta^0),\\
    c_9\left(\beta ^{-6 \epsilon -1}\right),c_{11}\left(\beta ^{2-6 \epsilon }\right),c_{12}(\beta^0),c_{12}\left(\beta ^{1-6 \epsilon }\right),c_{14}(\beta^0),c_{14}\left(\beta ^{-6 \epsilon -1}\right)\Big)^\intercal\,.
\end{multline}
The only nonzero constant is 
\begin{gather}
    c_1\left(\beta ^{9-10 \epsilon }\right)=-\frac{2 \pi ^{7/2} \csc (2 \pi  \epsilon ) \Gamma (2-2 \epsilon )}{\Gamma \left(\frac{11}{2}-5 \epsilon \right) \Gamma \left(\frac{3}{2}-\epsilon \right)^2 \Gamma (2 \epsilon -1)}\,.
\end{gather}
\subsection*{$(4,0)$ cuts}
\begin{figure}
    \centering
    \includegraphics[width=0.67\textwidth]{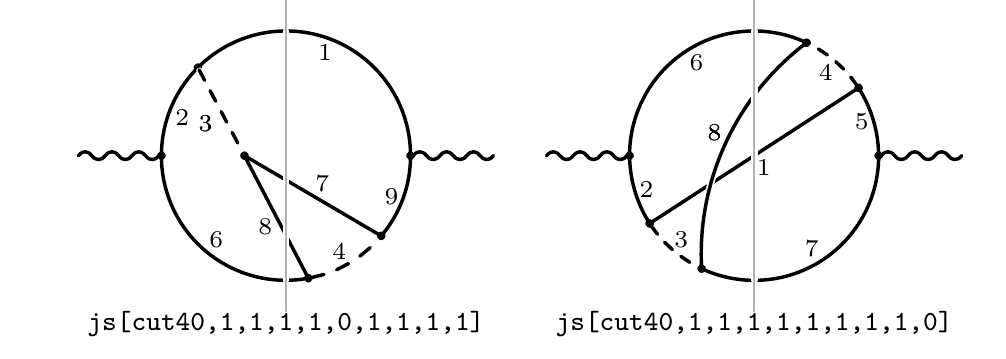}
    \caption{Prototypes for master integrals of $(4,0)$ cuts.}
    \label{fig:cut4m-protos}
\end{figure}

All Laporta master integrals for $(4,0)$ cuts belong to two prototypes in Fig. \ref{fig:cut4m-protos}. They are defined as
\begin{align}
    j^{\texttt{cut40}}_{n_1\ldots,n_9} &= \frac{(2\pi)^4}{2\left(\pi^{d/2}\right)^3}\int d^dl_1 d^dl_2 d^dl_3\frac{\prod_{k\in \{1,6,7,8\}}\delta^{(n_k-1)}(-D_k)}{\prod_{k=1,k\not\in \{1,6,7,8\}}^{9} (D_k-i0)^{n_k}}\,,
\end{align}
where one of $n_5, n_9$ is necessarily non-positive. We find 13 master integrals:
%\begin{gather}\nonumber
%    j_{100001110}^{\texttt{cut40}},\ 
%    j_{200001110}^{\texttt{cut40}},\ 
%    j_{300001110}^{\texttt{cut40}},\ 
%    j_{100001111}^{\texttt{cut40}},\ 
%    j_{101011110}^{\texttt{cut40}},\ 
%    j_{102011110}^{\texttt{cut40}},\
%    j_{101101110}^{\texttt{cut40}},\\
%    j_{201101110}^{\texttt{cut40}},\
%    j_{101102110}^{\texttt{cut40}},\
%    j_{110001111}^{\texttt{cut40}},\
%    j_{120001111}^{\texttt{cut40}},\
%    j_{111111110}^{\texttt{cut40}},\
%    j_{121111110}^{\texttt{cut40}}.
%\end{gather}

\begin{center}
    \includegraphics[width=\textwidth]{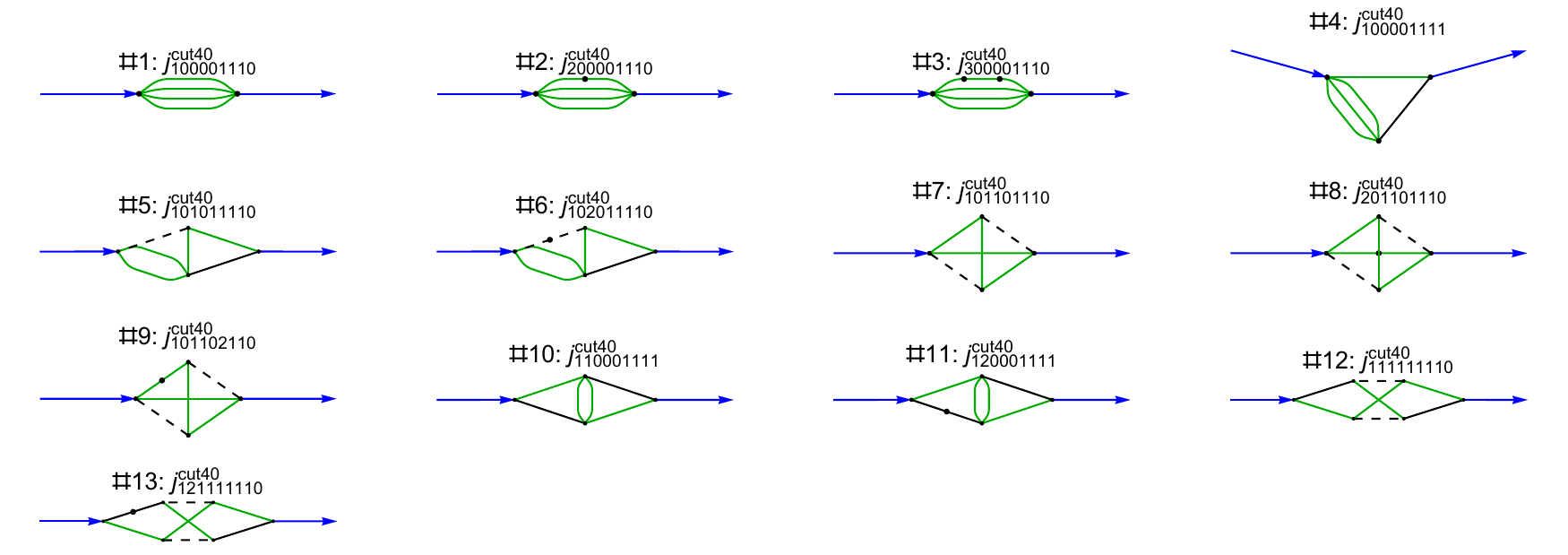}
\end{center}

Boundary conditions are fixed by the threshold asymptotics of the first integral (phase space of four massive particles).

\section{Results}
The results of our calculation in its full length are presented in the ancillary files. Let us describe briefly the content of these files:
\begin{description}
    \item \texttt{Data/weights} --- file with definition of weights. The function $f$ entering the some weights is defined in Eq. \eqref{eq:fdef} (see description of \texttt{Data/frules40} below). 
    \item \texttt{Data/jtoGraph$kl$} --- file containing graphs for the Laporta master integrals for the cut $(k,l)$.
    \item \texttt{Data/jtoJ$kl$} --- substitution rules for the Laporta master integrals in terms of the canonical/$\ep$-regular basis.
    \item \texttt{Data/M$kl$} --- matrix in the right-hand side of the differential system $\partial_s \bs J=M \bs J$.
    \item \texttt{Data/T$kl$} --- transformation matrix to $ep$-form/$ep$-regular basis.
    \item \texttt{cs$kl$} --- substitution rules for the coefficients of threshold asymptotic expansion used to fix the boundary conditions.
    \item \texttt{Data/JtoII$kl$} --- substitution rules for the canonical/$\ep$-regular basis in terms of the iterated integrals defined below.
    \item \texttt{Data/Jthr$kl$} --- substitution rules for the threshold asymptotics of the  canonical/$\ep$-regular basis. Note that the $(4,0)$ cut has threshold at $s=16$, in contrast to $(2,l)$ cuts.
    \item \texttt{Data/Jhigh$kl$} --- substitution rules for the high-energy asymptotics of the  canonical/$\ep$-regular basis.
    \item \texttt{Data/IItoG} --- file containing substitution rules for the iterated integrals in terms of Goncharov's polylogarithms.
    \item \texttt{Data/JtoN40} --- file containing the substitution rules for the numerical evaluation of the non-polylogarithmic master integrals for $(4,0)$ cut.
    \item \texttt{Data/IIrtoLeft40} --- ``unshuffling'' rules for moving $r_0,\ r_1,\ r_2,\ r_3$, and $\tilde{r}_3$ to  the left-most position.
    \item \texttt{Data/frules40} --- definition of function $f$, Eq. \eqref{eq:fdef}, and its differentiation rule.
    \item \texttt{masters-results.nb} --- this is the exemplary \textit{Mathematica} notebook showing the usage of the above data files.
\end{description}
Note that in the ancillary files we have chosen to present the $\ep$-expansions of the canonical/$\ep$-regular master integrals rather than those of Laporta master integrals. The reason is that it appears to be beneficial to first substitute in the cross section (or spectral density) the later via the former and then to substitute the expansions for canonical/$\ep$-regular integrals. Doing it this way one avoids unnecessary loss of the higher orders in $\ep$.

One more remark is in place here. In most physical applications the master integrals calculated in the present paper enter in the sum with their mirrored counterpart, which corresponds to complex conjugation. Therefore, only the real part of the results presented here is relevant for such applications (spectral density is one of them).

For expository and reference purposes we present the leading terms of $\ep$-expansion of the Laporta master integrals in the Appendix.

\section{Asymptotics}

Now let us describe in some details tha calculation of the threshold and and high energy asymptotics of the master integrals. The polylogarithmic master integrals for the $(2,0)$, $(2,1)$ and $(2,2)$ cuts were expressed in terms of multiple polylogarithms of argument $\beta$ and thus the calculation of their asymptotic expansions proceeds along the same lines. At threshold ($s=4$ or $\beta=0$) the expansion is particularly simply. All one needs to do is to expand integration kernels at small values of parameter and integrate the sum of the obtained monomials in integration variable. The corresponding results up to $\OO (\beta^5)$ can be fould in the  accompanying files \texttt{Jthr(2,*)}. To obtain the expansion in the high energy limit we first rewrite polylogarithms in $\beta$ in terms of polylogarithms of argument $1-\beta$ using recursively the following  transformation
\begin{equation}
G(a_1,\ldots ,a_k,\beta) = G(a_1,\ldots , a_k, 1) + \int_0^{1-\beta} \frac{dt}{t-(1-a_1)}G(a_2,\ldots ,a_k,\beta)\, .
\end{equation}  
To expand polylogarithms from $1-\beta$ in high energy limit we expand integration kernels and integration limit at large $s$. The obtained sum of monomials in integration variable are then easily integrated similar to the threshold case. The high-energy expansions up  $\OO (1/s^5)$ can be found in accompanying files \texttt{Jhigh(2,*)}. 

The asymptotic expansions for non-polylogarithmic master integrals for $(4,0)$ are somewhat different. First, we use shuffle relations to rewrite the expressions for master integrals in terms of iterated integrals with elliptic kernels in the first position. Then it is convenient to rewrite polylogarithmic integrals entering the last integration with elliptic kernel in terms of classical polylogarithms\footnote{For this purpose we use the \texttt{HPL} package \cite{HPL1,HPL2}.}. As a particular example let us consider the following element of $\ep$-regular basis:
\begin{equation}
    J_{06}^{\texttt{cut40}}(s) = \tfrac{1}{2} I\left(\left.l_2\tilde{r}_3\right|s\right)-I\left(\left.r_3\right|s\right)\,.
\end{equation} 
Using shuffle relations we move the transcendental weights $r_k$ to the left:
\begin{equation}
J_{06}^{\texttt{cut40}}(s) = -I(r_3|s) + \tfrac{1}{2}I(l_2|s) I(\tilde{r}_3|s) - \tfrac{1}{2}I(\tilde{r}_3,l_2|s)
\end{equation}
and rewrite the inner polylogarithmic integrals in terms of classical polylogarithms:
\begin{equation}
J_{06}^{\texttt{cut40}}(s) = -I(r_3|s) - \tfrac{1}{2}\ln\left(\tfrac{12}{s-4}\right)I(\tilde{r}_3|s) + \tfrac{1}{2}I\left(
\tilde{r}_3\ln\left(\tfrac{12}{s-4}\right)\Big|s
\right)\, .
\end{equation}
Now, we have a one-fold integral representation, whose threshold (at $s=16$) asymptotics can be easily found by Taylor expanding its integrand at $s=16$ and performing trivial integration. To find its high-energy asymptotics it is convenient to use the identity\footnote{Note that this identity is not literally applicable to most of the integrals under consideration because the first term in Eq. \eqref{eq:ehigh} diverges at large $s$. In this case we subtract from the integrand several terms of its high-energy asymptotics.}
\begin{equation}\label{eq:ehigh}
\int_{16}^s ds f(s) = \int_{16}^{\infty} ds f(s) - \int_s^{\infty} ds f(s)
\end{equation}   
The first term here is some constant which is difficult to determine by explicit analytical evaluation of the integral. Therefore, we use the numerical approach to recognize this constant as a rational combination of (alternating) multiple zeta values using PSLQ algorithm \cite{PSLQ}. In the second term we can safely perform Taylor expansion of the integrand at large $s$ and take the remaining integral term-wise. Note that this approach requires to recognize exactly one constant for each integral, independently on the chosen expansion order in $1/s$. The asymptotics of other master integrals can be computed along the same lines and the results can be found in the accompanying files \texttt{Jthr(4,0)} and \texttt{Jhigh(4,0)}.

We have checked that the obtained asymptotic expansions are consistent with numerical values of the corresponding integrals in the threshold and the high-energy regions.

\section{Conclusion}

In this paper we have presented the results for the master integrals entering the three-loop spectral density of photon self energy in QED. The non-polylogarithmic master integrals have been considered earlier in our work \cite{Lee:2019wwn}. Here we have calculated the remaining integrals which are expressible via Goncharov's polylogarithms. We used the differential equation method \cite{diffeqn1,diffeqn2,diffeqn3,diffeqn4,diffeqn5} and the possibility to reduce the corresponding differential systems to $\ep$-form \cite{epform1,epform2}. Given the available algorithms and tools, in particular, the recently published \textit{Mathematica} package \Libra \cite{lee2020libra}, the reduction to $\ep$-form appears to be very simple. We have chosen to fix the boundary conditions at the threshold point. We used \Libra to determine which asymptotic coefficients have to be calculated in order to fix the boundary conditions. A somewhat more involved step was the calculation of required coefficients using the expansion by regions method \cite{Beneke:1997zp}. We have managed to calculate all required coefficients exactly in $\ep$. This fact, given the found transformations to $\ep$ form, allows one to easily obtain even more terms of $\ep$ expansion if needed. To verify obtained expressions we used numerical checked with sector decomposition method \cite{Binoth:2000ps,Binoth:2003ak,Binoth:2004jv,Heinrich:2008si,Bogner:2007cr,Bogner:2008ry,Kaneko:2009qx} as implemented in \cite{Fiesta4}. 

To check results for master integrals with $(2,0)$ cuts the latter can be applied directly because of a simple relation of these integrals and the onshell vertex diagrams. In the case of $(2,1)$ and $(2,2)$ cuts the existing implementations can not be applied directly due to the presence of infrared divergences related to phase-space integration and necessity to perform the resolution of singularities also for these integration variables. This difficulty can be however avoided by considering integrals in sufficiently high dimensions \footnote{We use $d=6-2\ep$ and $d=8-2\ep$.}, where the integrals become infrared-finite. The original integrals are then obtained with the use of dimensional recurrence relations \cite{dimrecurrence}. Note, that in the case of $(2,2)$ cuts there are no ultraviolet divergent subgraphs and the sector decomposition is not required at all. In the case of $(2,1)$ cuts the corresponding divergence in parametric representation for one-loop subgraph is factorizes in terms of the external $\Gamma$-function factor while parametric integral itself is convergent. So, in this case sector decomposition is not required also. In addition, to check the results for master integrals with $(2,0)$ cuts we have reproduced the 2-loop vertex diagrams calculation as presented in \cite{Bonciani:2003te}. Finally, we have presented results for the threshold and high-energy asymptotics of the considered master integrals, including non-polylogarithmic master integrals with $(4,0)$ cuts, and checked their consistency with the obtained exact results.

\paragraph*{Acknowledgments} This work was supported by Russian Science Foundation, grant 20-12-00205.

\appendix

\section{Leading terms of $\ep$-expansion for Laporta master integrals}
We introduce the following weights
\begin{gather}\label{eq:weights1}
    l_0(s) = \frac{1}{s}, \quad l_1(s) = \frac{1}{s \beta}, \quad l_2(s) = \frac{1}{s -4}, 
    \quad l_4(s) = \frac{1}{s-1}\,,\\
    r_k(s) = \frac{f(s)}{s \beta^k}\theta(s-16)\quad (k=0,1,2,3),\\
    \tilde{r}_3(s) =\frac{8 \beta  (s+2) f(s)}{(s-16) (s-4)^2}\theta(s-16)\,,
\end{gather}
where 
\begin{gather}\label{eq:fdef}
    f(s)=\frac{16 (s-16)}{s}\left[
    \mathrm{K}(1-k_{-})\mathrm{K}(k_{+}) - \mathrm{K}(k_{-})\mathrm{K}(1-k_{+})
    \right],\\
    k_{\pm}=\frac12\left[1\pm\left(1-\frac8s\right) \sqrt{1-\frac{16}{s}}+\frac{16}{s} \sqrt{1-\frac{4}{s}}\right]\,.
\end{gather}
and the iterated integrals
\begin{equation}
    I(w_n,\ldots w_1|s)= \int\limits_{s>s_n>\ldots >s_1>4} \prod_{k=1}^{n} ds_k w_k(s_k)\,.
\end{equation}
\subsection*{$(2,0)$ cuts}
\begin{align}%(2,0)
    j_{000011101}^{\texttt{cut20}} &= \tfrac{\pi  \beta }{\epsilon ^2}+O(\tfrac{1}{\epsilon }),\\
    j_{001111100}^{\texttt{cut20}} &= -\tfrac{\pi  \beta }{2 \epsilon ^2}+O(\tfrac{1}{\epsilon }),\\
    j_{010011101}^{\texttt{cut20}} &= -\tfrac{\pi  \beta }{\epsilon ^2}+O(\tfrac{1}{\epsilon }),\\
    j_{010110110}^{\texttt{cut20}} &= -\tfrac{\pi  \beta }{\epsilon ^2}+O(\tfrac{1}{\epsilon }),\\
    j_{020110110}^{\texttt{cut20}} &= \tfrac{\pi  \beta }{2 \epsilon ^2}+O(\tfrac{1}{\epsilon }),\\
    j_{100011110}^{\texttt{cut20}} &= -\tfrac{3 (\pi  \beta )}{2 \epsilon ^2}+O(\tfrac{1}{\epsilon }),\\
    j_{011110101}^{\texttt{cut20}} &= \tfrac{\pi  \beta }{2 \epsilon ^2}+O(\tfrac{1}{\epsilon }),\\
    j_{011111100}^{\texttt{cut20}} &= \tfrac{\pi  \beta }{2 \epsilon ^2}+O(\tfrac{1}{\epsilon }),\\
    j_{101011110}^{\texttt{cut20}} &= \tfrac{\pi  \beta }{2 \epsilon ^2}+O(\tfrac{1}{\epsilon }),\\
    j_{201011110}^{\texttt{cut20}} &= \tfrac{2 \pi ^3 I(l_1|s)}{3 s}+\tfrac{2 i \pi ^2 I(l_1,l_1|s)}{s}-\tfrac{2 \pi  I(l_1,l_1,l_1|s)}{s}+O(\epsilon ^1),\\
    j_{110011101}^{\texttt{cut20}} &= \tfrac{\pi  \beta }{\epsilon ^2}+O(\tfrac{1}{\epsilon }),\\
    j_{110011110}^{\texttt{cut20}} &= \tfrac{\pi  \beta }{2 \epsilon ^2}+O(\tfrac{1}{\epsilon }),\\
    j_{210011110}^{\texttt{cut20}} &= \tfrac{\pi ^3 I(l_1|s)}{3 s}+\tfrac{i \pi ^2 I(l_1,l_1|s)}{s}-\tfrac{\pi  I(l_1,l_1,l_1|s)}{s}+O(\epsilon ^1),\\
    j_{011111101}^{\texttt{cut20}} &= \tfrac{i \pi  (2 \pi ^3+2 i \pi ^2 \ln\!{2}\,+21 i \zeta_3) I(l_1|s)}{2 s}+\tfrac{2 \pi ^3 I(l_0,l_1|s)}{3 s}-\tfrac{\pi ^3 I(l_1,l_0|s)}{2 s}+\tfrac{\pi ^3 I(l_1,l_1|s)}{s}-\tfrac{\pi ^3 I(l_1,l_2|s)}{s}\nonumber\\
    &+\tfrac{2 i \pi ^2 I(l_0,l_1,l_1|s)}{s}-\tfrac{i \pi ^2 I(l_1,l_0,l_1|s)}{s}+\tfrac{i \pi ^2 I(l_1,l_1,l_2|s)}{s}-\tfrac{2 i \pi ^2 I(l_1,l_2,l_1|s)}{s}-\tfrac{2 \pi  I(l_0,l_1,l_1,l_1|s)}{s}\nonumber\\
    &+\tfrac{\pi  I(l_1,l_0,l_1,l_1|s)}{s}-\tfrac{\pi  I(l_1,l_1,l_2,l_1|s)}{s}+\tfrac{2 \pi  I(l_1,l_2,l_1,l_1|s)}{s}+O(\epsilon ^1),\\
    j_{021111101}^{\texttt{cut20}} &= \tfrac1{\epsilon}\left(\tfrac{i \pi ^2 I(l_1|s) \beta }{(s-4) s}-\tfrac{\pi  I(l_1,l_1|s) \beta }{(s-4) s}+\tfrac{\pi ^3 \beta }{2 (s-4) s}\right)+O(\epsilon ^0),\\
    j_{111111110}^{\texttt{cut20}} &= \tfrac{1}{\epsilon }\Big(\tfrac{\pi ^2 \beta  (\pi +2 i \ln\!{2}\,) I(l_1|s)}{(s-4) s}+\tfrac{i \pi ^2 \beta  I(l_1,l_0|s)}{(s-4) s}+\tfrac{i \pi ^2 \beta  I(l_1,l_2|s)}{(s-4) s}-\tfrac{\pi  \beta  I(l_1,l_0,l_1|s)}{(s-4) s}-\tfrac{\pi  \beta  I(l_1,l_2,l_1|s)}{(s-4) s}\Big)\nonumber\\
    &+O(\epsilon ^0),\\
    j_{211111110}^{\texttt{cut20}} &= \tfrac{1}{\epsilon }\Big(
    \tfrac{i \pi ^2 I(l_0|s)}{2 s^2}+\tfrac{i \pi ^2 I(l_2|s)}{2 s^2}-\tfrac{\pi  I(l_0,l_1|s)}{2 s^2}+\tfrac{i \pi ^2 \beta  I(l_1,l_0|s)}{(s-4) s^2}+\tfrac{i \pi ^2 \beta  I(l_1,l_2|s)}{(s-4) s^2}-\tfrac{\pi  I(l_2,l_1|s)}{2 s^2}\nonumber\\
    &-\tfrac{\pi  \beta  I(l_1,l_0,l_1|s)}{(s-4) s^2}-\tfrac{\pi  \beta  I(l_1,l_2,l_1|s)}{(s-4) s^2}+\tfrac{\pi  (-\beta  s^2-4 \beta  s+4 i \pi  \ln\!{2}\, s+2 \pi ^2 s-8 i \pi  s-16 i \pi  \ln\!{2}\,-8 \pi ^2+16 i \pi )}{4 (s-4) s^2}\nonumber\\
    &+\tfrac{\pi  I(l_1|s) (2 s+\pi ^2 \beta +2 i \pi  \beta  \ln\!{2}\,-4)}{(s-4) s^2}
    \Big)+O(\epsilon ^0),\\
    j_{100011101}^{\texttt{cut20a}} &= -\tfrac{\pi  \beta }{\epsilon ^2}+O(\tfrac{1}{\epsilon }),\\
    j_{110011101}^{\texttt{cut20a}} &= \tfrac{\pi  \beta }{\epsilon ^2}+O(\tfrac{1}{\epsilon })
\end{align}
\subsection*{$(2,1)$ cuts}
\begin{align}%(2,1)
    j_{100100110}^{\texttt{cut21}} &= \tfrac{1}{\epsilon }\Big(\tfrac{2 \pi  (s-1) I(l_1|s)}{s}-\tfrac{1}{2} \pi  (s+2) \beta\Big)+O(\epsilon ^0),\\
    j_{200100110}^{\texttt{cut21}} &= \tfrac{1}{\epsilon }\Big(\pi  \beta -\tfrac{\pi  (s-2) I(l_1|s)}{s}\Big)+O(\epsilon ^0),\\
    j_{101101100}^{\texttt{cut21}} &= \tfrac{1}{\epsilon }\Big(\tfrac{1}{2} \pi  (s+2) \beta -\tfrac{2 \pi  (s-1) I(l_1|s)}{s}\Big)+O(\epsilon ^0),\\
    j_{102101100}^{\texttt{cut21}} &= \tfrac{1}{\epsilon }\Big(\tfrac{\pi  (s-2) I(l_1|s)}{s}-\pi  \beta \Big)+O(\epsilon ^0),\\
    j_{101102100}^{\texttt{cut21}} &= \pi  I(l_0|s) \beta +2 \pi  I(l_2|s) \beta +\pi  (-1-2 i \pi +2 \ln\!{2}\,) \beta +\tfrac{\pi  s I(l_0,l_1|s)}{2 (s-1)}\nonumber\\
    &-\tfrac{\pi  (s^2-2 s+4) I(l_1,l_0|s)}{2 (s-1) s}-\tfrac{\pi  (s^2-2 s+4) I(l_1,l_2|s)}{(s-1) s}\nonumber\\
    &+\tfrac{\pi  I(l_1|s) (-\ln\!{2}\, s^2+i \pi  s^2-2 s^2+2 \ln\!{2}\, s-2 i \pi  s+4 s-4 \ln\!{2}\,+4 i \pi -2)}{(s-1) s}+O(\epsilon ^1),\\
    j_{110100110}^{\texttt{cut21}} &= \tfrac{1}{\epsilon }\Big(\tfrac{1}{2} \pi  (s+2) \beta -\tfrac{2 \pi  (s-1) I(l_1|s)}{s}\Big)+O(\epsilon ^0),\\
    j_{120100110}^{\texttt{cut21}} &= -\tfrac{i \pi  (-i \beta  s+\pi  s-4 \pi )}{s}+i \pi ^2 I(l_0|s)+\tfrac{\pi  (s-2) I(l_1|s)}{s}-\pi  I(l_0,l_1|s)+O(\epsilon ^1),\\
    j_{110101100}^{\texttt{cut21}} &= \tfrac{1}{\epsilon }\Big(\tfrac{1}{2} \pi  (s+2) \beta -\tfrac{2 \pi  (s-1) I(l_1|s)}{s}\Big)+O(\epsilon ^0),\\
    j_{210101100}^{\texttt{cut21}} &= \tfrac{1}{\epsilon }\Big(\tfrac{\pi  (s-2) I(l_1|s)}{s}-\pi  \beta \Big)+O(\epsilon ^0),\\
    j_{101101110}^{\texttt{cut21}} &= -\tfrac{1}{2} \pi  I(l_0|s) \beta -\pi  I(l_2|s) \beta +\pi  (1+i \pi -\ln\!{2}\,) \beta +\tfrac{\pi  (s-2) I(l_1,l_0|s)}{2 s}+\tfrac{\pi  (s-2) I(l_1,l_2|s)}{s}\nonumber\\
    &+\pi  I(l_0,l_1,l_0|s)+2 \pi  I(l_0,l_1,l_2|s)+\tfrac{\pi  (s-1) I(l_4,l_0,l_1|s)}{2 s}-\tfrac{3 \pi  (s-1) I(l_4,l_1,l_0|s)}{2 s}\nonumber\\
    &-\tfrac{3 \pi  (s-1) I(l_4,l_1,l_2|s)}{s}+\tfrac{\pi  I(l_1|s) (2 \ln\!{2}\, s-2 i \pi  s+s-4 \ln\!{2}\,+4 i \pi )}{2 s}\nonumber\\
    &+\tfrac{1}{2} \pi  I(l_0,l_1|s) (-1-4 i \pi +4 \ln\!{2}\,)+\tfrac{3 i \pi  (s-1) I(l_4,l_1|s) (\pi +i \ln\!{2}\,)}{s}+O(\epsilon ^1),\\
    j_{101102110}^{\texttt{cut21}} &= \pi  \beta -\tfrac{\pi  (s-2) I(l_1|s)}{s}+O(\epsilon ^1),\\
    j_{201101110}^{\texttt{cut21}} &= \tfrac{3 i \pi  (\pi +i \ln\!{2}\,) I(l_4,l_1|s)}{s}+\tfrac{\pi  I(l_4,l_0,l_1|s)}{2 s}-\tfrac{3 \pi  I(l_4,l_1,l_0|s)}{2 s}-\tfrac{3 \pi  I(l_4,l_1,l_2|s)}{s}+O(\epsilon ^1),\\
    j_{101111100}^{\texttt{cut21}} &= \tfrac{1}{\epsilon }\Big(\pi  \beta -\tfrac{\pi  (s-2) I(l_1|s)}{s}\Big)+O(\epsilon ^0),\\
    j_{110100111}^{\texttt{cut21}} &= \tfrac{1}{\epsilon }\Big(\pi  \beta -\tfrac{\pi  (s-2) I(l_1|s)}{s}\Big)+O(\epsilon ^0),\\
    j_{120100111}^{\texttt{cut21}} &= \tfrac{2 \pi  I(l_1,l_1,l_1|s)}{s}-\tfrac{i \pi ^2 I(l_1,l_1|s)}{s}+O(\epsilon ^1),\\
    j_{110101110}^{\texttt{cut21}} &= \tfrac{\pi  (\beta  s+i \pi  s-4 i \pi )}{s}-\tfrac{\pi  (s-2) I(l_1|s)}{s}-\tfrac{2 i \pi ^2 I(l_1,l_1|s)}{s}+\tfrac{4 \pi  I(l_1,l_1,l_1|s)}{s}+O(\epsilon ^1),\\
    j_{110102110}^{\texttt{cut21}} &= -\tfrac{i \pi ^2 (s-4) I(l_0|s)}{s}+\tfrac{\pi  (i \pi  \beta  s^2-6 i \pi  \beta  s+2 s+4 i \pi  \beta -8) I(l_1|s)}{(s-4) s}+\tfrac{\pi  (s-4) I(l_0,l_1|s)}{s}\nonumber\\
    &-\tfrac{2 \pi  (s^2-6 s+4) \beta  I(l_1,l_1|s)}{(s-4) s}+\tfrac{\pi  (s-4) I(l_2,l_1|s)}{s}-\tfrac{2 i \pi  (-i \beta  s+\pi  \ln\!{2}\, s-4 \pi  \ln\!{2}\,-\pi )}{s}+O(\epsilon ^1),\\
    j_{111101100}^{\texttt{cut21}} &= -\tfrac{\pi }{2}  I(l_0|s) \beta -\pi  I(l_2|s) \beta +\pi  (1+i \pi -\ln\!{2}\,) \beta +\tfrac{1}{2} \pi  I(l_0,l_1|s)+\tfrac{\pi  (s-2) I(l_1,l_0|s)}{2 s}\nonumber\\
    &-\tfrac{2 i \pi  (2 \pi -i s \beta ) I(l_1,l_1|s)}{s}+\tfrac{\pi  (s-2) I(l_1,l_2|s)}{s}+\tfrac{8 \pi  I(l_1,l_1,l_1|s)}{s}+\tfrac{\pi  I(l_4,l_0,l_1|s)}{2 s}-\tfrac{3 \pi  I(l_4,l_1,l_0|s)}{2 s}\nonumber\\
    &-\tfrac{3 \pi  I(l_4,l_1,l_2|s)}{s}+\tfrac{\pi  I(l_1|s) (2 i \pi  \beta  s+2 \ln\!{2}\, s-2 i \pi  s+s-4 \ln\!{2}\,+4 i \pi )}{2 s}+\tfrac{3 i \pi  I(l_4,l_1|s) (\pi +i \ln\!{2}\,)}{s}+O(\epsilon ^1),\\
    j_{111111100}^{\texttt{cut21}} &= \tfrac{2 i \pi ^2 I(l_0,l_1,l_1|s)}{s}-\tfrac{i \pi ^2 I(l_1,l_0,l_1|s)}{s}-\tfrac{2 i \pi ^2 I(l_1,l_2,l_1|s)}{s}-\tfrac{4 \pi  I(l_0,l_1,l_1,l_1|s)}{s}-\tfrac{\pi  I(l_0,l_4,l_0,l_1|s)}{2 s}\nonumber\\
    &+\tfrac{3 \pi  I(l_0,l_4,l_1,l_0|s)}{2 s}+\tfrac{3 \pi  I(l_0,l_4,l_1,l_2|s)}{s}+\tfrac{2 \pi  I(l_1,l_0,l_1,l_1|s)}{s}-\tfrac{2 \pi  I(l_1,l_1,l_1,l_0|s)}{s}-\tfrac{4 \pi  I(l_1,l_1,l_1,l_2|s)}{s}\nonumber\\
    &+\tfrac{4 \pi  I(l_1,l_2,l_1,l_1|s)}{s}+\tfrac{3 \pi  I(l_0,l_4,l_1|s) (-i \pi +\ln\!{2}\,)}{s}+\tfrac{4 i \pi  I(l_1,l_1,l_1|s) (\pi +i \ln\!{2}\,)}{s}+O(\epsilon ^1),\\
    j_{121111100}^{\texttt{cut21}} &= \tfrac{\pi  I(l_1|s)}{4 s \epsilon ^2}+O(\tfrac{1}{\epsilon }),\\
    j_{111111110}^{\texttt{cut21}} &= \tfrac{1}{\epsilon ^2}\Big(\tfrac{i \pi ^2 \beta  I(l_1|s)}{2 (s-4) s}-\tfrac{\pi  \beta  I(l_1,l_1|s)}{(s-4) s}\Big)+O(\tfrac{1}{\epsilon }),\\
    j_{121111110}^{\texttt{cut21}} &= \tfrac{1}{\epsilon ^2}\Big(\tfrac{i \pi  (i s \beta +2 \pi )}{4 (s-4) s^2}+\tfrac{\pi  (-i \pi  \beta  s-2 s+2 i \pi  \beta +8) I(l_1|s)}{2 (s-4)^2 s^2}+\tfrac{\pi  (s-2) \beta  I(l_1,l_1|s)}{(s-4)^2 s^2}\Big)+O(\tfrac{1}{\epsilon })
\end{align}
\subsection*{$(2,2)$ cuts}
\begin{align}%(2,2)
    j_{101100010}^{\texttt{cut22}} &= \tfrac{1}{12} \pi  (s^2+20 s+12) \beta -\tfrac{\pi  (s-1) (s+2) I(l_1|s)}{s}+2 \pi  I(l_0,l_1|s)+O(\epsilon ^1),\\
    j_{201100010}^{\texttt{cut22}} &= -\tfrac{1}{4} \pi  (5 s+6) \beta +\tfrac{\pi  (s^2+4 s-6) I(l_1|s)}{2 s}-2 \pi  I(l_0,l_1|s)+O(\epsilon ^1),\\
    j_{102100010}^{\texttt{cut22}} &= \tfrac{1}{\epsilon }\Big(\tfrac{2 \pi  (s-1) I(l_1|s)}{s}-\tfrac{1}{2} \pi  (s+2) \beta \Big)+O(\epsilon ^0),\\
    j_{101100011}^{\texttt{cut22}} &= -\tfrac{1}{4} \pi  (s-6) \beta +2 \pi  I(l_1,l_1|s) \beta -\tfrac{\pi  (s-3) I(l_1|s)}{s}-\pi  I(l_0,l_1|s)+O(\epsilon ^1),\\
    j_{101101110}^{\texttt{cut22}} &= -2 \pi  I(l_1,l_1|s) \beta -\pi  \beta +\tfrac{\pi  (s-2) I(l_1|s)}{s}+\pi  I(l_0,l_1|s)-\pi  I(l_0,l_0,l_1|s)\nonumber\\
    &+\tfrac{2 \pi  (s-2) I(l_1,l_1,l_1|s)}{s}+O(\epsilon ^1),\\
    j_{101102110}^{\texttt{cut22}} &= \tfrac{1}{\epsilon }\Big(\tfrac{\pi  (s-2) I(l_1|s)}{2 s}-\tfrac{\pi  \beta }{2}\Big)+O(\epsilon ^0),\\
    j_{201101110}^{\texttt{cut22}} &= \tfrac{4 \pi  I(l_1,l_1,l_1|s)}{s}+O(\epsilon ^1),\\
    j_{111100011}^{\texttt{cut22}} &= -\pi  \beta +\tfrac{\pi  (s-2) I(l_1|s)}{s}-\tfrac{4 \pi  I(l_1,l_1,l_1|s)}{s}+O(\epsilon ^1),\\
    j_{111111110}^{\texttt{cut22}} &= \tfrac{2 \pi  \beta  I(l_1,l_1|s)}{(s-4) s \epsilon ^2}+O(\tfrac{1}{\epsilon }),\\
    j_{121111110}^{\texttt{cut22}} &= \tfrac{1}{\epsilon ^2}\Big(-\tfrac{2 \pi  (s-2) I(l_1,l_1|s) \beta }{(s-4)^2 s^2}+\tfrac{\pi  \beta }{2 (s-4) s}+\tfrac{2 \pi  I(l_1|s)}{(s-4) s^2}\Big)+O(\tfrac{1}{\epsilon }),\\
    j_{011111100}^{\texttt{cut22a}} &= (-2 \pi  I(l_1,l_1|s) \beta -\pi  \beta +\tfrac{2 \pi  I(l_1|s)}{s}+\tfrac{\pi  (s-2) I(l_0,l_1|s)}{s}+\tfrac{\pi  (s-4) I(l_2,l_1|s)}{s})+O(\epsilon ^1),\\
    j_{011111110}^{\texttt{cut22a}} &= (\tfrac{2 \pi  I(l_0,l_1,l_1,l_1|s)}{s}-\tfrac{\pi  I(l_1,l_1,l_0,l_1|s)}{s}-\tfrac{\pi  I(l_1,l_1,l_2,l_1|s)}{s})+O(\epsilon ^1),\\
    j_{021111110}^{\texttt{cut22a}} &= -\tfrac{\pi  \beta  I(l_1,l_1|s)}{(s-4) s \epsilon }+O(\epsilon ^0),\\
    j_{111111110}^{\texttt{cut22a}} &= \tfrac{\pi  \beta  I(l_1,l_1|s)}{(s-4) s \epsilon ^2}+O(\tfrac{1}{\epsilon }),\\
    j_{211111110}^{\texttt{cut22a}} &= \tfrac{1}{\epsilon ^2}\Big(\tfrac{\pi  I(l_1,l_1|s) \beta }{(s-4) s^2}-\tfrac{\pi  \beta }{2 (s-4) s}+\tfrac{\pi  I(l_1|s)}{2 s^2}\Big)+O(\tfrac{1}{\epsilon })
\end{align}
\subsection*{$(4,0)$ cuts}
\begin{align}%(4,0)
    j_{100001110}^{\texttt{cut40}} &= \tfrac{\pi  s (s^2+82 s-128) f(s)}{24 (s-16)^2}+\tfrac{\pi  (s^4-22 s^3-1152 s^2+2432 s+4096) f'(s)}{288 (s-16)}\nonumber\\
    &+\tfrac{1}{288} \pi  (s-4) s (s^2+40 s+64) f''(s)+O(\epsilon ^1),\\
    j_{200001110}^{\texttt{cut40}} &= -\tfrac{9 \pi  f(s) s^2}{8 (s-16)^2}-\tfrac{1}{96} \pi  (s-4) (5 s+16) f''(s) s\nonumber\\
    &-\tfrac{\pi  (s+2) (3 s^2-176 s+512) f'(s)}{96 (s-16)}+O(\epsilon ^1),\\
    j_{300001110}^{\texttt{cut40}} &= \tfrac{\pi  (s^2+18 s-64) f(s)}{8 (s-16)^2}+\tfrac{\pi  (s+2) (s^2-44 s+128) f'(s)}{96 (s-16)}\nonumber\\
    &+\tfrac{1}{96} \pi  (s-4) s (s+4) f''(s)+O(\epsilon ^1),\\
    j_{100001111}^{\texttt{cut40}} &= -\tfrac{1}{96} \pi  (s^2-26 s+64) f'(s)-\tfrac{1}{96} \pi  (s-16) (s-4) s f''(s)+O(\epsilon ^1),\\
    j_{101011110}^{\texttt{cut40}} &= -\tfrac{\pi  I(r_2|s)}{6 s}-\tfrac{1}{24} \pi  \beta  I(\tilde{r}_3|s)+\tfrac{\pi  I(l_1,\tilde{r}_3|s)}{12 s}-\tfrac{\pi  (5 s-32) f(s)}{4 (s-16)^2}+\tfrac{\pi  (11 s-32) f'(s)}{12 (s-16)}\nonumber\\
    &-\tfrac{1}{24} \pi  (s-4) s f''(s)+O(\epsilon ^1),\\
    j_{102011110}^{\texttt{cut40}} &= -\tfrac{1}{48} \pi  \beta  I(\tilde{r}_3|s)+\tfrac{\pi  (5 s+64) f(s)}{8 (s-16)^2}+\tfrac{\pi  (3 s^2-122 s+320) f'(s)}{48 (s-16)}\nonumber\\
    &+\tfrac{1}{16} \pi  (s-4) s f''(s)+O(\epsilon ^1),\\
    j_{101101110}^{\texttt{cut40}} &= -\tfrac{1}{24} \pi  I(r_0|s)+\tfrac{\pi  (s-2) I(r_2|s)}{12 s}+\tfrac{1}{24} \pi  \beta  I(\tilde{r}_3|s)-\tfrac{\pi  (s-2) I(l_1,\tilde{r}_3|s)}{24 s}-\tfrac{\pi  (s^2-2 s+64) f(s)}{24 (s-16)^2}\nonumber\\
    &+\tfrac{\pi  (11 s-32) f'(s)}{12 (s-16)}-\tfrac{1}{24} \pi  (s-4) s f''(s)+O(\epsilon ^1),\\
    j_{201101110}^{\texttt{cut40}} &= \tfrac{\pi  I(r_2|s)}{6 s}-\tfrac{\pi  I(l_1,\tilde{r}_3|s)}{12 s}+O(\epsilon ^1),\\
    j_{101102110}^{\texttt{cut40}} &= -\tfrac{1}{24} \pi  \beta  I(\tilde{r}_3|s)-\tfrac{\pi  (5 s-32) f(s)}{4 (s-16)^2}+\tfrac{\pi  (11 s-32) f'(s)}{12 (s-16)}-\tfrac{1}{24} \pi  (s-4) s f''(s) +O(\epsilon ^1),\\
    j_{110001111}^{\texttt{cut40}} &= -\tfrac{\pi  I(r_2|s)}{12 s}+\tfrac{1}{48} \pi  \beta  I(\tilde{r}_3|s)+\tfrac{\pi  I(l_1,\tilde{r}_3|s)}{24 s}-\tfrac{\pi  (s+8) f(s)}{2 (s-16)^2}-\tfrac{\pi  (s^2-42 s+128) f'(s)}{24 (s-16)}\nonumber\\
    &-\tfrac{1}{24} \pi  (s-4) s f''(s)+O(\epsilon ^1),\\
    j_{120001111}^{\texttt{cut40}} &= \tfrac{\pi  (3 s-8) \beta  I(\tilde{r}_3|s)}{48 (s-4) s}+\tfrac{\pi  (5 s-8) f(s)}{6 (s-16) (s-4) s}-\tfrac{1}{24} \pi  f'(s)+O(\epsilon ^1),\\
    j_{111111110}^{\texttt{cut40}} &= -\tfrac{\pi  \beta  I(l_1,r_0|s)}{24 (s-4) s}+\tfrac{\pi  \beta  I(l_1,r_2|s)}{6 (s-4) s}-\tfrac{\pi  \beta  I(l_1,l_1,\tilde{r}_3|s)}{12 (s-4) s}+O(\epsilon ^1),\\
    j_{121111110}^{\texttt{cut40}} &= (-\tfrac{\pi  I(r_0|s)}{24 (s-4) s^2}+\tfrac{\pi  \beta  I(r_1|s)}{8 (s-4)^2 s}+\tfrac{\pi  I(r_2|s)}{6 (s-4) s^2}-\tfrac{\pi  (s-16) \beta  I(\tilde{r}_3|s)}{96 (s-4) s^2}-\tfrac{\pi  \beta  I(l_0,\tilde{r}_3|s)}{8 (s-4)^2 s}+\tfrac{\pi  (s-2) \beta  I(l_1,r_0|s)}{24 (s-4)^2 s^2}\nonumber\\
    &-\tfrac{\pi  (s-2) \beta  I(l_1,r_2|s)}{6 (s-4)^2 s^2}-\tfrac{\pi  I(l_1,\tilde{r}_3|s)}{12 (s-4) s^2}+\tfrac{\pi  (s-2) \beta  I(l_1,l_1,\tilde{r}_3|s)}{12 (s-4)^2 s^2})+O(\epsilon ^1)
\end{align}

\bibliographystyle{JHEP}
\bibliography{PO}

\end{document}